\documentclass[journal]{IEEEtran}
\makeatletter

\usepackage{blindtext}
\usepackage[T1]{fontenc}
\usepackage{textcomp}
\usepackage{amsmath}
\usepackage{amssymb}
\usepackage{bm}
\usepackage{mdwmath}
\usepackage{cases}
\usepackage[short, c2]{optidef}
\usepackage{graphicx}
\graphicspath{{Figure/PDF/}{Biography/PDF/}}
\usepackage[dvipsnames]{xcolor}
\definecolor{myblue}{rgb}{0,0.4980,1} 
\definecolor{myred}{rgb}{0.8706,0.1608,0.0627} 
\usepackage[caption=false,font=footnotesize,subrefformat=parens]{subfig}
\usepackage[nosort,noadjust]{cite}
\usepackage{url}
\usepackage{tabularray}
\UseTblrLibrary{varwidth}
\UseTblrLibrary{diagbox}
\ExplSyntaxOn
\tl_const:Nn \c__tblr_valign_M_tl {M}
\tl_new:N    \g__tblr_cell_valign_sub_tl

\keys_define:nn { tblr-column }
  {
    M .meta:n = { valign = M },
  }

\cs_gset_protected:Npn \__tblr_get_cell_alignments:nn #1 #2
  {
    \group_begin:
    \tl_gset:Nx \g__tblr_cell_halign_tl
      { \__tblr_data_item:neen { cell } {#1} {#2} { halign } }
    \tl_set:Nx \l__tblr_v_tl
      { \__tblr_data_item:neen { cell } {#1} {#2} { valign } }
    \tl_case:NnF \l__tblr_v_tl
      {
        \c__tblr_valign_t_tl
          {
            \tl_gset:Nn \g__tblr_cell_valign_tl {m}
            \tl_gset:Nn \g__tblr_cell_middle_tl {t}
            \tl_gclear:N \g__tblr_cell_valign_sub_tl
          }
        \c__tblr_valign_m_tl
          {
            \tl_gset:Nn \g__tblr_cell_valign_tl {m}
            \tl_gset:Nn \g__tblr_cell_middle_tl {m}
            \tl_gclear:N \g__tblr_cell_valign_sub_tl
          }
        \c__tblr_valign_M_tl
          {
            \tl_gset:Nn \g__tblr_cell_valign_tl {m}
            \tl_gset:Nn \g__tblr_cell_valign_sub_tl {M}
            \tl_gset:Nn \g__tblr_cell_middle_tl {t}
          }
        \c__tblr_valign_b_tl
          {
            \tl_gset:Nn \g__tblr_cell_valign_tl {m}
            \tl_gset:Nn \g__tblr_cell_middle_tl {b}
            \tl_gclear:N \g__tblr_cell_valign_sub_tl
          }
      }
      {
        \tl_gset_eq:NN \g__tblr_cell_valign_tl \l__tblr_v_tl
        \tl_gclear:N \g__tblr_cell_middle_tl
      }
    \group_end:
  }

\cs_gset_protected:Npn \__tblr_build_cell_content:NN #1 #2
  {
    \hbox_set_to_wd:Nnn \l__tblr_a_box { \l__tblr_cell_wd_dim }
      {
        \tl_if_eq:NnTF \g__tblr_cell_halign_tl {j}
          { \__tblr_get_cell_text:nn {#1} {#2} \hfil }
          {
            \tl_if_eq:NnF \g__tblr_cell_halign_tl {l} { \hfil }
            \__tblr_get_cell_text:nn {#1} {#2}
            \tl_if_eq:NnF \g__tblr_cell_halign_tl {r} { \hfil }
          }
      }
    \vbox_set_to_ht:Nnn \l__tblr_b_box { \l__tblr_cell_ht_dim }
      {
        \tl_case:Nn \g__tblr_cell_valign_tl
          {
            \c__tblr_valign_m_tl
              {
                \vfil
                \bool_lazy_and:nnT
                  { \int_compare_p:nNn { \l__tblr_cell_rowspan_tl } < {2} }
                  { !\tl_if_eq_p:NN \c__tblr_valign_M_tl \g__tblr_cell_valign_sub_tl }
                  {
                    \box_set_ht:Nn \l__tblr_a_box
                      { \__tblr_data_item:nen { row } {#1} { @row-upper } }
                    \box_set_dp:Nn \l__tblr_a_box
                      { \__tblr_data_item:nen { row } {#1} { @row-lower } }
                  }
                \box_use:N \l__tblr_a_box
                \vfil
              }
            \c__tblr_valign_h_tl
              {
                \box_set_ht:Nn \l__tblr_a_box
                  { \__tblr_data_item:nen { row } {#1} { @row-head } }
                \box_use:N \l__tblr_a_box
                \vfil
              }
            \c__tblr_valign_f_tl
              {
                \vfil
                \int_compare:nNnTF { \l__tblr_cell_rowspan_tl } < {2}
                  {
                    \box_set_dp:Nn \l__tblr_a_box
                      { \__tblr_data_item:nen { row } {#1} { @row-foot } }
                  }
                  {
                    \box_set_dp:Nn \l__tblr_a_box
                      {
                        \__tblr_data_item:nen
                          { row }
                          { \int_eval:n { #1 + \l__tblr_cell_rowspan_tl - 1 } }
                          { @row-foot }
                      }
                  }
                \box_use:N \l__tblr_a_box
              }
          }
        \hrule height ~ 0pt 
      }
    \vbox_set_to_ht:Nnn \l__tblr_c_box
      { \l__tblr_row_ht_dim - \l__tblr_row_abovesep_dim }
      {
        \box_use:N \l__tblr_b_box
        \vss
      }
    \skip_horizontal:n { \l__tblr_x_tl }
    \box_use:N \l__tblr_c_box
    \skip_horizontal:n { \l__tblr_y_tl - \l__tblr_cell_wd_dim + \l__tblr_w_tl }
  }
\ExplSyntaxOff
\usepackage{pifont}

\usepackage{hyperref}
\newcommand{\colorhypersetup}{\@ifpackageloaded{hyperref}{\hypersetup{%
	bookmarksopen=true,%
	bookmarksnumbered=true,%
	pdfpagemode={UseOutlines},
	pdfstartview={FitH},%
	colorlinks=true,%
	linkcolor={myred},%
	citecolor={orange}
}}{\empty}}
\newcommand{\blackhypersetup}{\@ifpackageloaded{hyperref}{\hypersetup{%
	bookmarksopen=true,%
	bookmarksnumbered=true,%
	pdfpagemode={UseOutlines},
	pdfstartview={FitH},%
	colorlinks=true,%
	allcolors={black}
}}{\empty}}
 \blackhypersetup

\usepackage{mfirstuc-english}
\MFUhyphentrue
\usepackage[version3,description,IEEEtran]{eacro}
\acsetup{%
    pages/display=all,%
    pages/seq/use=false,%
    pages/name=true,%
    pages/fill={\quad},%
    make-links=false%
}
\DeclareAcronym{udt}{
	short = UDT,
	long = user DT}
 \DeclareAcronym{dt}{
	short = DT,
	long = digital twin}
 \DeclareAcronym{idt}{
	short = IDT,
	long = infrastructure DT}
 \DeclareAcronym{sdt}{
	short = SDT,
	long = slice DT}
 \DeclareAcronym{nc}{
	short = NC,
	long = network controller}
 \DeclareAcronym{ap}{
        short = AP,
        long  = access point    
 }
  \DeclareAcronym{gai}{
        short = GAI,
        long  = generative \ac{ai}    
 }
  \DeclareAcronym{gan}{
        short = GAN,
        long  = generative adversarial network    
 }
  \DeclareAcronym{gt}{
        short = GT,
        long  = generative transformer    
 }
 \DeclareAcronym{vae}{
        short = VAE,
        long  = variational autoencoder    
 }
 \DeclareAcronym{gdm}{
        short = GDM,
        long  = generative diffusion model   
 }
  \DeclareAcronym{ai}{
        short = AI,
        long  = artificial intelligence   
 }
  \DeclareAcronym{gdt}{
        short = GDT,
        long  = GAI-driven DT
 }
   \DeclareAcronym{gail}{
        short = GAIL,
        long  = generative adversarial imitation learning
 }
    \DeclareAcronym{drl}{
        short = DRL,
        long  = deep reinforcement learning
 }
    \DeclareAcronym{mdrl}{
        short = MDRL,
        long  = model-based DRL
 }
  \DeclareAcronym{iot}{
        short = IoT,
        long  = Internet of Things
 }
   \DeclareAcronym{qoe}{
        short = QoE,
        long  = quality of experience
 }
    \DeclareAcronym{qos}{
        short = QoS,
        long  = quality of service
 }
    \DeclareAcronym{mg}{
        short = MG,
        long  = multicast group
 }
    \DeclareAcronym{smg}{
        short = SMG,
        long  = sub-MG
 }

\usepackage{algpseudocode}
\newcounter{MYalgorithmic}

{%
	\vspace*{3pt}
	\hrule height 1pt}
\newcounter{MYitem}[MYalgorithmic]

\newcommand{\MYlabel}[1]{\def\@currentlabel{\theALG@line}\label{#1}}

\newcommand{\upperroman}[1]{\uppercase\expandafter{\romannumeral#1}}

\newcommand{\myunit}[1]{%
	\ifmmode
		\mathrm{#1}
	\else
		$ \mathrm{#1} $
	\fi}
\newcommand{\murm}{%
	\ifmmode
		\text{\textmu}
	\else
		\textmu
	\fi}

\newcommand{\MYnewpage}{%
	\ifCLASSOPTIONonecolumn
		\ifCLASSOPTIONjournal
			\typeout{The onecolumn journal mode.}
			\newpage
		\fi
	\fi}

\newlength{\mysinglefigwidth}
\newlength{\mymultifigwidth}
\ifCLASSOPTIONonecolumn
	\AtBeginDocument{\setlength{\mysinglefigwidth}{0.7\linewidth}}
\else
	\AtBeginDocument{\setlength{\mysinglefigwidth}{\linewidth}}
\fi
\ifCLASSOPTIONonecolumn
	\AtBeginDocument{\setlength{\mymultifigwidth}{0.5\linewidth}}
\else
	\AtBeginDocument{\setlength{\mymultifigwidth}{\linewidth}}
\fi

\makeatother

\begin{document}
\title{When Digital Twin Meets Generative AI: Intelligent Closed-Loop Network Management}
\author{Xinyu Huang,~\IEEEmembership{Student Member,~IEEE}, Haojun Yang,~\IEEEmembership{Member,~IEEE}, Conghao~Zhou,~\IEEEmembership{Member, IEEE}, Mingcheng~He,~\IEEEmembership{Student Member, IEEE}, Xuemin~(Sherman)~Shen,~\IEEEmembership{Fellow,~IEEE}, Weihua~Zhuang,~\IEEEmembership{Fellow,~IEEE}
    \thanks{Xinyu Huang, Haojun Yang, Conghao Zhou, Mingcheng He, Xuemin (Sherman) Shen, and Weihua Zhuang are with the Department of Electrical and Computer Engineering, University of Waterloo, Waterloo, ON, N2L 3G1, Canada (E-mail: \{x357huan, haojun.yang, c89zhou, m64he, sshen, wzhuang\}@uwaterloo.ca).}
}

\ifCLASSOPTIONonecolumn
	\typeout{The onecolumn mode.}
\else
	\typeout{The twocolumn mode.}
\fi

\maketitle

\ifCLASSOPTIONonecolumn
	\typeout{The onecolumn mode.}
	\vspace*{-50pt}
\else
	\typeout{The twocolumn mode.}
\fi
\begin{abstract}
Generative artificial intelligence (GAI) and digital twin (DT) are advanced data processing and virtualization technologies to revolutionize communication networks. Thanks to the powerful data processing capabilities of GAI, integrating it into DT is a potential approach to construct an intelligent holistic virtualized network for better network management performance. To this end, we propose a GAI-driven DT (GDT) network architecture to enable intelligent closed-loop network management. In the architecture, various GAI models can empower DT status emulation, feature abstraction, and network decision-making. The interaction between GAI-based and model-based data processing can facilitate intelligent external and internal closed-loop network management. To further enhance network management performance, three potential approaches are proposed, i.e., model light-weighting, adaptive model selection, and data-model-driven network management. We present a case study pertaining to data-model-driven network management for the GDT network, followed by some open research issues.
\end{abstract}

\ifCLASSOPTIONonecolumn
	\typeout{The onecolumn mode.}
	\vspace*{-10pt}
\else
	\typeout{The twocolumn mode.}
\fi

\IEEEpeerreviewmaketitle

\MYnewpage


\section{Introduction}
\label{sec:Introduction}
\IEEEPARstart{O}{ver} the past decade, \ac{dt} technology has emerged as a key virtualization technique, experiencing significant growth and development. {\ac{dt} was first introduced to monitor and mitigate anomalous events for flying vehicles via accurately simulating their entire lifecycles}~\cite{glaessgen2012digital}. {The mobile communication network usually consists of the core network and the radio access network, which are mainly responsible for data routing and wireless transmission, respectively.} {By applying \ac{dt} in mobile communication networks, holistic network virtualization for efficient network management can be realized~\cite{9839640,holi}.} {Specifically, \ac{dt} consists of three modules to support efficient network management, i.e., 1) network status emulation module to mirror the status of physical mobile communication networks through advanced prediction-based algorithms, which can reduce frequent data collection cost; 2) {data feature abstraction module to distill the useful patterns of network traffic and user behaviors}, which can simplify network management problems; 3) network decision-making module to make tailored network management strategies, which can be validated in the virtualized environment.} Since the above modules require efficient and accurate data processing in the intricate and dynamic network environment, the utilization of advanced data processing techniques is imperative.

As an emerging branch of \ac{ai}, \ac{gai} focuses on generating new data instances, analyzing data correlation, and addressing optimization problems~\cite{10398474}, which can help improve network status emulation, data feature abstraction, and network decision-making in \ac{dt}. For instance, \ac{gan} can generate high-fidelity network scenario images and expand datasets through its generator and discriminator~\cite{goodfellow2020generative}, {which can empower \ac{dt} emulation module.} \Ac{gt} performs excellently in text understanding and generation due to its attention mechanism~\cite{vaswani2017attention}, which can assist \ac{dt} in perceiving user intentions and analyzing data correlation. Furthermore, \ac{gdm} can facilitate conditional generation and decision-making by its state diffusion and inverse dynamics mechanism~\cite{ho2020denoising}, thereby improving \ac{dt} network decision-making performance. {By integrating \ac{gai} and \ac{gt} into mobile communication networks, {an intelligent \ac{gdt} network architecture for external and internal closed-loop network management can be realized.} Specifically, the external loop emphasizes the interaction between the physical network and \ac{dt}, where the data quality of generated \ac{dt} status via \ac{gai} is evaluated by an error discriminator in \ac{dt} for adaptive data collection frequency. {The internal loop focuses on the interaction between \ac{dt} and \ac{gai} in the \ac{gdt}, where the abstracted features are separately fed into the model-based network management module in \ac{dt} and the \ac{gai}-based network management module to evaluate which one can achieve best network performance for adaptive network management policy adjustment.}

{However, achieving intelligent external and internal closed-loop network management for the \ac{gdt} network architecture poses technical challenges, including}
\begin{itemize}
    \item \textit{Massive caching and computing overhead:} {Since GAI models usually have large model sizes and complex neural network structures, it is challenging to directly deploy GAI models on network edge nodes with limited caching and computing capabilities.}
    \item \textit{Model scalability and efficiency:} {For the \ac{gai} models with different sizes, the larger \ac{gai} models usually provide better data processing performance but also consume more network resources. Therefore, how to select an appropriate model size with enough efficiency to adapt to network dynamics is challenging.}
    \item \textit{{Reliable data processing for network robustness:}} {Since the complexity and “black box” nature of GAI models may lead to unpredictable and biased status emulation, feature abstraction, and network decision-making in DT, it is challenging to design a reliable data processing mechanism to improve network robustness.}
\end{itemize}

\begin{table*}[t]
\renewcommand{\IEEEiedlistdecl}{\setlength{\IEEElabelindent}{0pt}}
\centering
\caption{Part of Emerging GAI Models Applied to Mobile Communication Networks}
\label{gai-a}
\begin{tblr}{
    width = \linewidth,
    colspec = {X[0.4,c,M]X[0.9,l,m]X[1.0,l,m]X[0.5,c,m]X[0.7,c,m]X[1.1,l,m]},
    hlines,
    hline{2} = {1}{-}{},
    hline{2} = {2}{-}{},
    vline{2-6},
    row{1} = {c,font=\bfseries},
    column{1} = {font=\bfseries},
    columns = {rightsep=3pt},
    cell{3}{4} = {l},
    cell{2}{5} = {c},
    measure=vbox,
}
Model & Functions & Characteristics & Storage Requirement & Computing Requirement & Metrics \\ 
GAN & \begin{itemize}
    \item Image generation
    \item Dataset augmentation
\end{itemize} & \begin{itemize}
    \item Generator
    \item Discriminator
\end{itemize} & $3\sim25~\myunit{GB}$ & $0.2\sim1~\myunit{PFlops}$ & \begin{itemize}
    \item Peak signal-to-noise ratio
    \item Structural similarity
\end{itemize} \\ 
GT & \begin{itemize}
    \item Text understanding \& generation
    \item Feature extraction
\end{itemize} & \begin{itemize}
    \item Positional encoding
    \item Self \& multi-head attention
    \item Residual connection
\end{itemize} & $10\sim30~\myunit{GB}$ & $0.5\sim2~\myunit{PFlops}$ &  \begin{itemize}
    \item General language understanding evaluation
    \item Perplexity
\end{itemize} \\ 
VAE & \begin{itemize}
    \item Feature learning
    \item Denoising
\end{itemize} & \begin{itemize}
    \item Probabilistic graphic model
    \item Variational inference
\end{itemize} & $20\sim800~\myunit{MB}$ & $1\sim100~\myunit{TFlops}$ & \begin{itemize}
    \item Mean squared error
    \item Kullback-Leibler divergence
\end{itemize} \\ 
GDM & \begin{itemize}
    \item Conditional generation
    \item Decision-making
\end{itemize} & \begin{itemize}
    \item Reverse diffusion
    \item Markovian transition
\end{itemize} & $0.3\sim3~\myunit{GB}$ & $2\sim200~\myunit{TFlops}$ & \begin{itemize}
    \item Temporal coherence
    \item Interpolation quality
\end{itemize} \\ 
\end{tblr}
\end{table*}

{In this article, {we propose a novel \ac{gdt} network architecture to realize intelligent external and internal closed-loop network management.} {Four kinds of typical \ac{gai} models, i.e., \ac{gan}, \ac{gt}, \ac{vae}, and \ac{gdm}, are selected to assist \ac{dt} status emulation, feature abstraction, and network decision-making.} The intelligent closed-loop network management includes two aspects. Specifically, {the interaction between \ac{gai}-based status emulation and \ac{dt}-based error discriminator can facilitate adaptive external closed-loop data collection.} The interplay between \ac{gai}-based and model-based decision-making can enable adaptive internal closed-loop network decision-making. {To address the aforementioned challenges, we first propose a model light-weighting method to reduce model caching and computing overhead. Secondly, we develop an adaptive model selection mechanism to adapt to network dynamics. {Thirdly, we design a data-model-driven method to improve network management robustness.} A case study pertaining to data-model-driven network management for \ac{gdt} is presented, followed by a discussion on potential research issues.}}

The remainder of this article is organized as follows. Firstly, \ac{dt} and advanced \ac{gai} techniques are discussed, followed by the proposed GDT network architecture. {Then, we discuss the challenges for GDT network and some potential solutions. Next, a case study about data-model-driven network management for \ac{gdt} network is presented.} Finally, the open research issues are identified, followed by the conclusion.

\section{Generative AI-Driven Digital Twin Network Architecture}

In this section, we first introduce \ac{dt} and advanced \ac{gai} techniques, and then propose a GDT network architecture.

\subsection{Digital Twin}
\subsubsection{Definition}
{DT is a virtual representation, also termed “black box”, of the physical mobile communication network, that reflects the real-time network status and provides efficient management strategies through a variety of embedded data-based models. These models can enable high-fidelity network status emulation, accurate network feature abstraction, and tailored network management strategies to facilitate efficient closed-loop network management.}
\subsubsection{Composition and Functionality}
{DT mainly consists of three modules, i.e., status emulation module, data feature abstraction module, and network decision-making module.}
\begin{itemize}
    \item {Status emulation module is the basis of DT, which is utilized to characterize the real-time status of physical mobile communication network. {The input of status emulation module is the collected data from the physical network through \acp{ap} and sensors}, which can be classified into networking-related data and behavior-related data. For example, the networking-related data includes users' channel conditions and service delay, base stations' transmission capabilities and traffic load, cloud and edge servers' caching and computing workload, etc. The behavior-related data includes users' interaction frequency, locations, mobility speed, preferences, etc. The status emulation module relies on the prediction-based algorithms, such as long short-term memory (LSTM) and recurrent neural network (RNN), to predict future network status. The output of status emulation module is the emulated network status, which can provide holistic network information for network management.}
    \item {Data feature abstraction module is responsible for distilling useful information from network status. The input of data feature abstraction module includes two aspects, i.e., emulated and realistic network status. The advanced data processing algorithms in DT are responsible for abstracting data features. For instance, the autoencoder can compress the high-dimensional data (time-series channel conditions, locations, and traffic load, etc.) into a low-dimensional latent representation. The output of data feature abstraction module is distilled network information, such as spatiotemporal traffic distribution and swipe probability distribution, {which can help the network controller capture the network dynamics.}}
    \item {Network decision-making module is responsible for outputting tailored network management decisions. The input of network decision-making module includes network status and abstracted data features. Network decision-making module is responsible for outputting tailored network management decisions. The input of network decision-making module includes network status and abstracted data features. To provide efficient network decisions, the data-based methods in DT can outperform traditional model-based methods. For instance, policy gradient methods can optimize network management policies in an online and incremental way, which can adapt to network dynamics. The output of network decision-making module is the network management decision, which is transferred to the network controller for implementation in the mobile communication network.}
\end{itemize} 
\subsubsection{Benefits}

{By introducing DT into mobile communication networks, three benefits can be obtained. Firstly, the real-time status of physical entity is essential for the network controller to make accurate resource management decisions, but can bring massive signaling overhead. The status emulation module in DT can emulate the real-time status of physical entities to reduce the traffic load. Secondly, the network status information often exhibits certain levels of data contamination and missing. If network resource management is conducted directly based on these data, it could degrade network performance. The data abstraction module in DT can provide correct and distilled information to the network controller. Thirdly, due to the diversity of users' service demands, such as some users prioritizing service latency and others focusing on the quality of transmitted content, DT can analyze their differences in service requirements to make tailored resource management.}

\subsection{Generative AI}

\Ac{gai} refers to a subset of \ac{ai} technologies that can autonomously generate novel content, data, and solutions. \ac{gai} has developed rapidly due to advancements in machine learning algorithms and the increasing availability of large datasets for training. {A key characteristic of \ac{gai} is its ability to learn intrinsic characteristics in the data, which can generate new data that closely mimic the original data distribution}~\cite{vaswani2017attention}. Therefore, {integrating \ac{gai} into mobile communication networks can generate high-fidelity network status and user behaviors, accurate data features, and optimized network configurations}~\cite{10398474}. As shown in Table~\ref{gai-a}, we select four representative \ac{gai} models that can be well applied to mobile communication networks for performance enhancement.

\begin{itemize}
    \item Generative adversarial network (GAN): It consists of a dueling generator and discriminator that refine generated data in a unique adversarial process~\cite{goodfellow2020generative}. By generating high-quality network scenario images and expanding datasets, \ac{gan} can improve the efficiency of virtual scene construction and network status emulation in \ac{dt}.
    \item Generative transformer (GT): It excels in text understanding and generation as well as feature abstraction through advanced positional encoding, self and multi-head attention mechanisms, and residual connections among neural networks~\cite{vaswani2017attention}. {For instance, a novel multi-modal mutual attention-based sentiment analysis framework was proposed to process complicated contexts and mine the association between unique semantics and common semantics~\cite{10124248}. {\ac{gt} can assist \ac{dt} in accurately perceiving user intentions and analyzing data correlations, thereby providing tailored network management information.}}
    \item Variational autoencoder (VAE): It is pivotal for feature abstraction and data recovery in mobile communication networks through probabilistic graphic models and variational inference~\cite{DBLP}. Specifically, {it can compress high-dimensional networking data into a low-dimensional latent representation and reconstruct data based on the captured data distribution.}
    \item Generative diffusion model (GDM): It facilitates conditional generation and network decision-making through reverse diffusion and Markovian transition processes~\cite{ho2020denoising}. {By modeling the network management process as a return-conditional diffusion model, the training efficiency and inference performance can be enhanced.}
\end{itemize}

By leveraging powerful status generation, data analytics, and decision-making capabilities, \ac{gai} can effectively enhance \ac{dt} capabilities.

\subsection{Generative AI-Driven DT Network Architecture}

\begin{figure}[!t]
    \centering
    \includegraphics[width=\mysinglefigwidth]{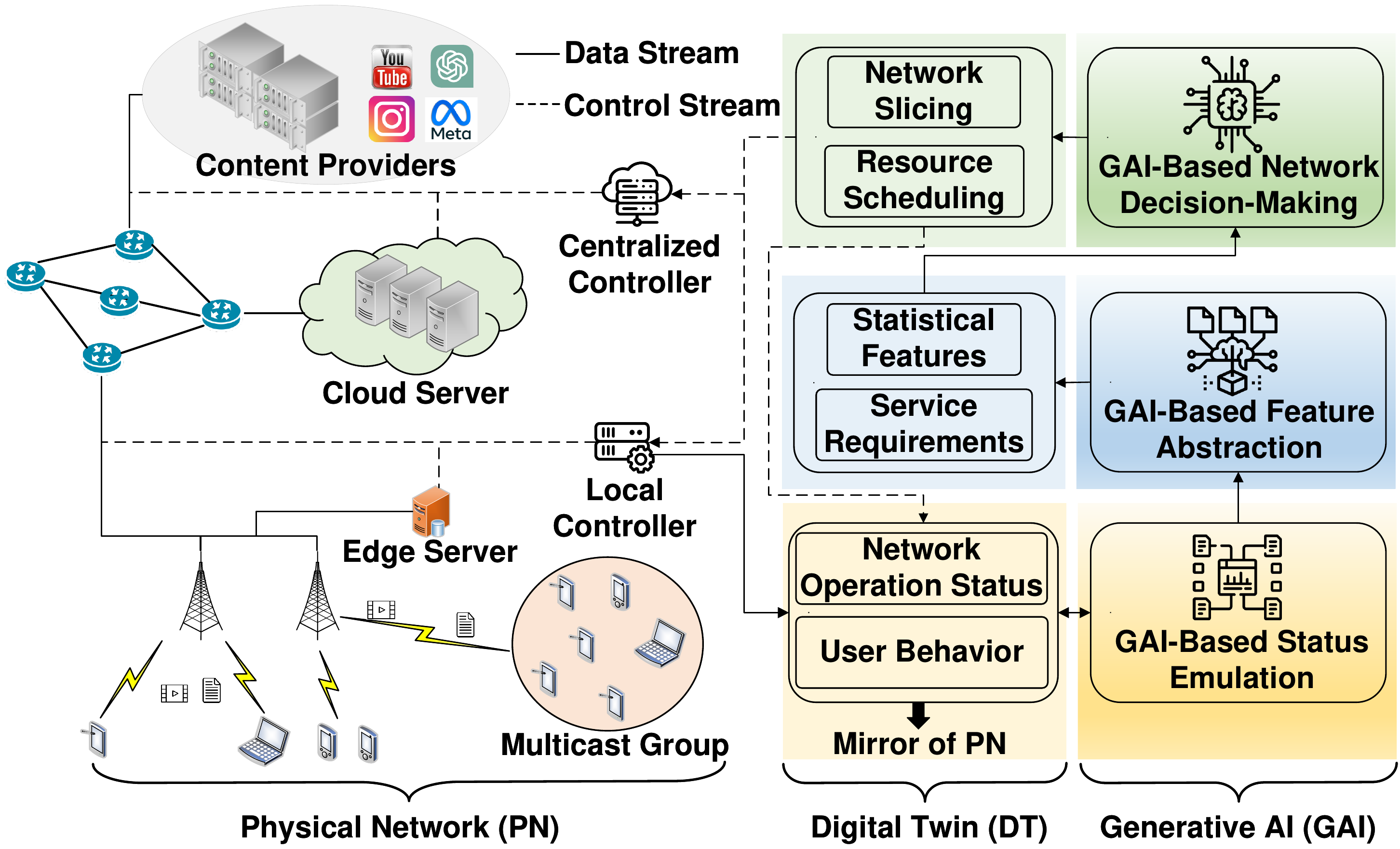}
    \caption{{The GDT network architecture.}}
    \label{network}
\end{figure}

\begin{figure*}[!t]
    \centering
    \includegraphics[width=1.8\mysinglefigwidth]{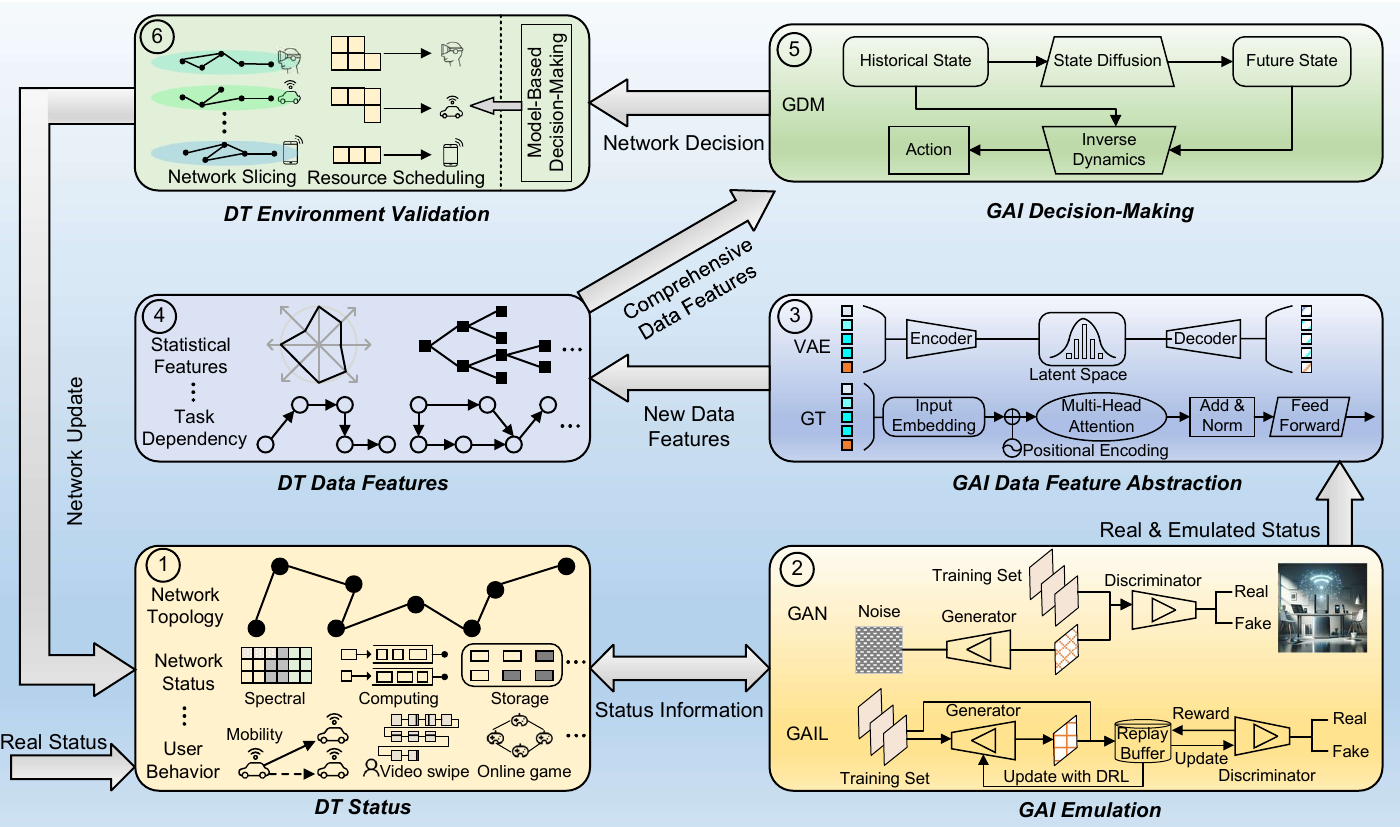}
    \caption{{The specific module interaction in GDT part.}}
    \label{Framework}
\end{figure*}

{To seamlessly integrate \ac{dt} and \ac{gai} in mobile communication networks, as shown in Fig.~\ref{network}, we develop a \ac{gdt} network architecture, which can effectively improve network management performance.} The physical network includes real-world network infrastructures, such as cloud and edge servers for data caching and computing, as well as \acp{ap} for unicast, multicast, and broadcast transmissions. {The virtual network consists of \ac{dt} and \ac{gai}, where they highly collaborate in the network status emulation, data feature abstraction, and network decision-making to improve network management performance.} {The local controller is responsible for small-timescale data collection and resource scheduling policy implementation, {while the centralized controller implements the large-timescale network slicing, i.e., the partitioning of a single physical network infrastructure into multiple and isolated logical networks~\cite{huang2023}.}}
{The specific module interaction in the \ac{gdt} part consists of \ac{gai}-based status emulation, feature abstraction, and decision-making, as shown in Fig.~\ref{Framework}.}

\subsubsection{GAI-Based Status Emulation}
{The boxes \ding{172} and \ding{173} in Fig.~\ref{Framework} show the status emulation process in \ac{gdt} part. Specifically, {\ac{dt} status consists of networking-related data (users’ channel conditions and service delay, base stations’ transmission capabilities and traffic load, cloud and edge servers’ caching and computing workload, etc.) and behavior-related data (users’ interaction frequency, locations, mobility speed, preferences, etc.), to mirror the real-time physical network status.} DT data needs to be periodically updated to guarantee freshness. {To reduce the frequent data collection overhead, advanced \ac{gai} techniques are utilized to generate high-fidelity \ac{dt} status.} For instance, \ac{gan} can help generate realistic network scenes for network environment construction and update by collaboratively training its generator and discriminator. Furthermore, \ac{gail} integrates the status generation capability of \ac{gan} into \ac{drl} algorithms, which can accurately emulate users' dynamic behaviors~\cite{bhattacharyya2022modeling}. Through real-time interaction between \ac{dt} status and \ac{gai} emulation, {high-fidelity \ac{dt} status can be effectively supplemented.}}

\subsubsection{GAI-Based Feature Abstraction}
Based on the realistic and emulated DT status, GAI-based feature abstraction is conducted in the boxes \ding{174} and \ding{175} in Fig.~\ref{Framework}. {Specifically, since \ac{dt} status data is usually high-dimensional and time-series, such as varying channel conditions and users' locations in a high-density network, it is hard to directly use them to guide efficient network management. Therefore, accurate and efficient data feature abstraction is necessary. For instance, {\ac{vae} can encode the high-dimensional DT status data into a low-dimensional representation to capture the network traffic patterns and user behavioral patterns.} Furthermore, \ac{gt} can accurately analyze spatiotemporal correlations from DT status through its advanced attention mechanism. The abstracted features can simplify the network management problem and provide tailored network management information.}

\subsubsection{GAI-Based Decision-Making}
The new and previous DT data features are integrated to generate comprehensive inputs for GAI-based network decision-making, as shown in boxes \ding{176} and \ding{177} in Fig.~\ref{Framework}. {Since the network management problem is usually complex and non-convex, it is difficult to directly use model-based or data-based methods to solve the problem to obtain a near-optimal solution. Therefore, advanced decision-making algorithms are necessary. For instance, {\ac{gdm} utilizes the state diffusion method to generate the future state (network status) based on the learned data distribution.} The generated state is integrated with the historical state as an input to the inverse dynamics mechanism for action (network management decision) generation. This can achieve a better convergence performance compared with \ac{drl} algorithms~\cite{ajay2022conditional}. {The generated network management decision is fed back to the DT environment for validation and DT status update.}}

\subsubsection{Procedure of Closed-Loop Network Management}
\begin{figure}[t]
    \centering
    \includegraphics[width=\mysinglefigwidth]{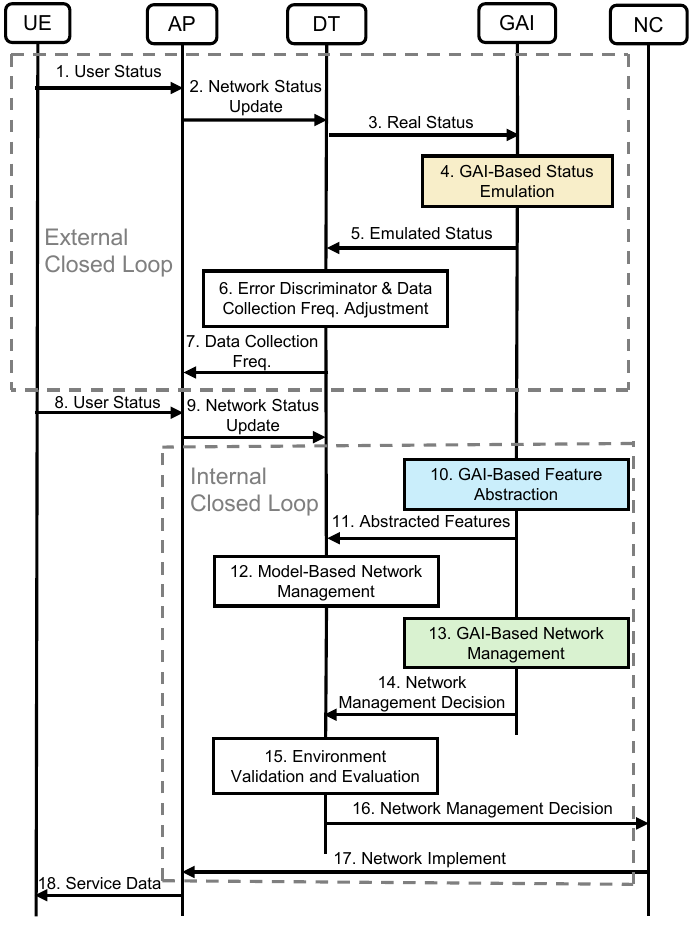}
    \caption{{Procedure of external and internal closed-loop network management.}}
    \label{workflow}
\end{figure}

As shown in Fig.~\ref{workflow}, intelligent external and internal closed-loop network management is realized in the \ac{gdt} network. {\textit{In the external closed loop, an adaptive data collection frequency mechanism is realized.} Specifically, user status is first uploaded to \ac{ap} to update \ac{dt} status with a prescribed data collection frequency. Then, the updated \ac{dt} status is sampled to a mini-batch that is transferred to \ac{gai} to conduct status emulation. Next, {the error discriminator in \ac{dt} evaluates the quality of newly generated status that is utilized to adjust data collection frequency.} Finally, the adjusted data collection frequency is fed back to the \ac{ap} to implement a new round of data collection.} {\textit{In the internal closed loop, an adaptive network decision-making mechanism is realized.}} Specifically, the realistic and emulated \ac{dt} status is first transferred to \ac{gai} to abstract data features. Then, the abstracted features are input to the model-based network management module in \ac{dt} and the \ac{gai}-based network management module, respectively. Next, {the generated network management decisions from two modules are validated in \ac{dt} environment to evaluate which one can achieve better network performance.} Finally, the better network management decision is fed back to the network controller for practical implementation. {Through the intelligent external and internal closed-loop network management, network performance can be effectively enhanced.}

\subsubsection{Advantages} {By integrating GAI with DT, four main advantages can be obtained. First, real-time network status and user behaviors can be accurately emulated by GAI algorithms without frequent data interaction between the physical network and DT, which can effectively reduce data collection cost. Then, the advanced GAI algorithms can compress high-dimensional DT data and analyze categorical features, which can simplify the network management problem. Next, GAI algorithms can solve complex network management problems and validate the solution in the constructed DT environment for further adjustment. Finally, the closed loop between GAI and DT can facilitate an adaptive network management decision, which can effectively improve network performance.}

\section{Challenges and Solutions}
To realize the proposed intelligent closed-loop network management in the \ac{gdt} network architecture, {optimizing \ac{gai} model deployment, model selection, and data processing in the GDT network is necessary, with some key research challenges.}

\subsection{Challenges}
\subsubsection{Massive Caching and Computing Overhead}
From the caching standpoint, the large \ac{gai} model size poses a challenge for effective model storage and retrieval. {Traditional caching strategies may not be sufficient especially in the network edge nodes, as the voluminous model parameters ranging from millions to billions exceed conventional caching capacities. Therefore, it is hard to directly deploy \ac{gai} models on network edge nodes with limited caching capabilities.} Moreover, {the large sizes and complex neural network structures of GAI models demand substantial computing resources}, which leads to increased latency and higher energy consumption. Therefore, it is challenging to improve the computing efficiency of \ac{gai} models for computation-intensive data processing.

\subsubsection{Model Scalability and Efficiency}

\ac{gai} models usually have various sizes, with larger models typically offering better data processing performance for network management but also requiring more caching and computing resources, whereas smaller models have the opposite characteristics. {To efficiently utilize a \ac{gai} model under varying network dynamics, an adaptive model selection mechanism is necessary, which can achieve a balance between model scalability and efficiency. For instance, a network might experience varying levels of congestion throughout the day, which affects service latency. A small \ac{gai} model that performs well during low-traffic periods might struggle when the network is congested, resulting in inconsistent service quality and degraded user experience. {Therefore, how to adaptively select an appropriate \ac{gai} model size with enough efficiency for network management presents a significant challenge.}}

\subsubsection{{Reliable Data Processing for Network Robustness}}

The utilization of \ac{gai} models in \ac{dt} may introduce instability due to several inherent drawbacks. The complexity and ``black box'' nature of \ac{gai} models can lead to unpredictable and biased decisions. Specifically, if \ac{gai} models fail to accurately grasp the operational mechanisms of physical networks, or cannot conduct thorough inferences under conditions of limited network resources, then the emulated network status and abstracted features may be biased. Moreover, {the network decision-making process is usually a complex optimization problem}, which requires professional theoretics to transform the problem for the contraction of the feasible solution set. Solely relying on the data training without adding any professional optimization theoretics may lead to a sub-optimal solution. {These factors combined underscore the challenge of reliable data processing for robust network management.}
\subsection{Solutions}
\subsubsection{Model Light-Weighting}

To handle the challenge of massive caching and computing overhead, {model split and knowledge distillation are effective methods to deploy lightweight \ac{gai} models.} Specifically, model split involves dividing a large \ac{gai} model into smaller and more manageable segments for distributed model caching and computing in the mobile communication networks~\cite{10040976}. Through efficient model split, massive data computing can be processed in parallel, {which can effectively reduce model inference latency and satisfy caching constraints.} {Knowledge distillation offers a complementary approach by transferring the knowledge from a large cumbersome teacher model to a small student model without significant performance loss}~\cite{Park_2019_CVPR}. Since the distilled student model is significantly less resource-intensive, it can be flexibly deployed within mobile communication networks with low caching and computing overhead. Based on the above analysis, the massive caching and computing overhead can be effectively mitigated through model split and knowledge distillation.
\subsubsection{Adaptive Model Selection}
{To deal with the challenge of \ac{gai} model scalability and efficiency, an adaptive \ac{gai} model selection mechanism is necessary. Specifically, {the \ac{gai} model selection mechanism needs to quantitatively assess the impact of network status, such as bandwidth availability and computing queue congestion}, as well as \ac{gai} model characteristics including caching and computing demands. {By employing a machine learning-based classification method, the \ac{gai} model selection mechanism can select the most suitable \ac{gai} model size for any given network scenario based on historical performance data, such as \ac{qos} and \ac{qoe}.} The \ac{gai} model selection mechanism also includes a feedback loop mechanism, where real-time performance data are used to continuously refine the classification algorithms and ensure that \ac{gai} models remain accurate and effective in the face of evolving network dynamics and service diversity. {By integrating these elements, an adaptive \ac{gai} model selection mechanism can be developed.}}
\subsubsection{Data-Model-Driven Network Management}
{To handle the challenge of reliable data processing for network robustness, data-model-driven methods are effective due to the holistic integration of empirical data and theoretical modeling.} {Unlike data-based methods that rely solely on machine learning algorithms to process historical data}, possibly overlooking underlying physical network principles, {data-model-driven methods incorporate the principles through mathematical modeling}, which can ensure a more comprehensive understanding of network dynamics and user behavior patterns. Based on the data-model-driven methods, the status emulation and feature abstraction can be more accurate and robust. Furthermore, in the face of complex network decision-making problems, {appropriate problem decomposition allows for an effective data-model-driven solution. Specifically, data-based methods usually excel in identifying state transition probability to solve a part of decoupled subproblems}, while model-based methods rely on classical optimization techniques to obtain the optimal solution to the remaining decoupled subproblems. Through the cross-iteration between data-based and model-based methods, a near-optimal and robust solution to the original network decision-making problem can be obtained.

\section{Case Study: Data-Model-Driven Network Management for GDT Network}

{In this section, a case study is provided on data-model-driven network management for \ac{gdt} network, aimed at improving \ac{qoe}.}

\subsection{Considered Scenario}

A GDT-assisted multicast short video streaming scenario is considered, which consists of two \acp{ap}, multiple \acp{mg}, and one GDT. In each scheduling slot, {bandwidth and computing resources are allocated to each \ac{smg} to receive videos with adaptive bitrate with the objective of maximizing QoE. The GDT consists of a status emulation module, a data feature abstraction module, and a network decision-making module, i.e.,}
\begin{itemize}
\item {In the GDT status emulation module, we first generate users' trajectories within the University of Waterloo campus with the Levy flight model that refers to a random Markovian walk and the probability distribution of step lengths satisfies heavy-tailed distribution. The generated data is used as the label data, where eighty percent of label data is selected as the training sample to train the LSTM model for users' trajectory emulation. The well-trained LSTM model is used to emulate users' trajectories. Based on users' real-time trajectories, the real-time channel conditions are emulated based on \texttt{PropagationModel} at Matlab by analyzing the channel fading between users and base stations. Users' swipe behaviors and preferences are sampled from the real-world video swipe dataset. Through the GDT status emulation module, users' real-time status can be perceived on the network side.}

\item {In the GDT data feature abstraction module, we propose an improved GAI method to update multicast groups and abstract users' swipe probability distribution. Specifically, the improved GAI method consists of three parts, i.e., autoencoder, double deep Q-network (DDQN), and K-means++. The autoencoder is selected as the basic GAI model that can reduce user status dimension by analyzing temporal correlation. Double deep Q-network is further used to mine the compressed user status to find an appropriate clustering number, while K-means++ is responsible for a fast and accurate multicast group update based on a given clustering number and compressed user status. In each updated multicast group, the swipe probability distribution is calculated based on users' swipe frequencies on different types of videos. Through the GDT data feature abstraction module, users' intrinsic features can be obtained for tailored network decision-making.}

\item {In the GDT network decision-making module, a data-model-driven method is proposed to realize multicast segment buffering and resource scheduling. Specifically, in the multicast segment buffering, we first utilize the abstracted swipe probability distribution from the GDT data feature abstraction module to analyze the watching probability of segments. Based on the analyzed watching probability and estimated multicast transmission capabilities, the multicast segment buffering number and order are determined. Then, the branching dueling Q-network is used to determine the segment version selection for each \ac{smg}. In the resource scheduling, the sequential least squares quadratic programming method is utilized to determine the occupation time of network resources by each \ac{smg}. The joint multicast segment buffering and resource scheduling decision is fed back to the DT environment to obtain the system reward, i.e., QoE, for Q-network update. Through the proposed data-model-driven method, complex network decision-making problems can be efficiently solved. The detailed simulation setting of network decision-making module can be found in~\cite{huang2024digital}.}
\end{itemize}

\subsection{Simulation Results}
\begin{figure}[t]
    \centering
    \includegraphics[width=0.9\mysinglefigwidth]{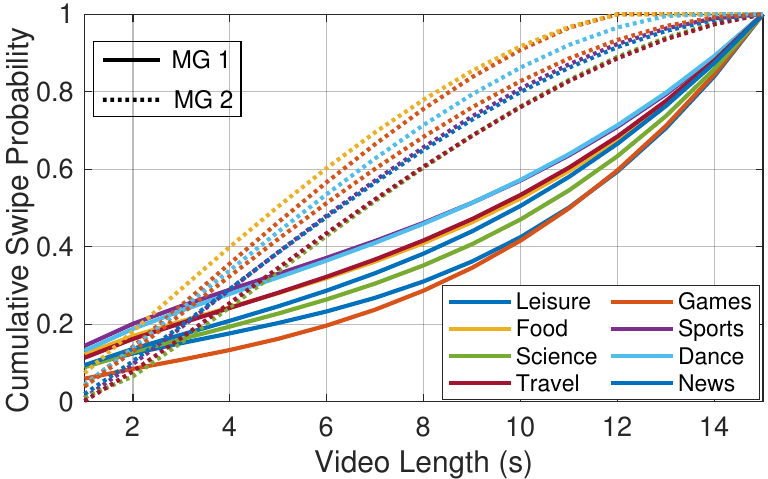}
    \caption{Cumulative swipe probability abstracted by the improved autoencoder.}
    \label{swipe}
\end{figure}

\begin{figure}[t]
    \centering
    \includegraphics[width=0.9\mysinglefigwidth]{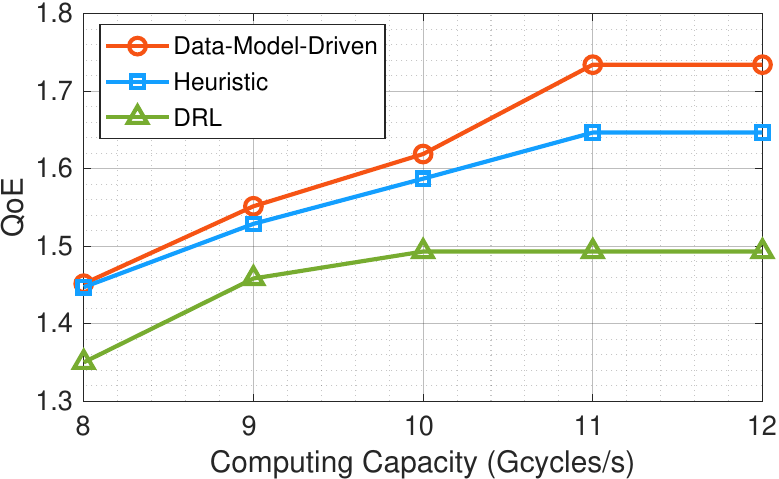}
    \caption{QoE vs. different computing capacities.}
    \label{VM}
\end{figure}

As illustrated in Fig.~\ref{swipe}, we present the cumulative swipe probability abstracted via \ac{gai}, where different line styles and colors represent various \acp{mg} and video types, respectively. It can be observed that users in \ac{mg} 1 have similar cumulative swipe probability as those in \ac{mg} 2 during the initial phase, but a noticeable divergence ensues over time. Since \ac{dt} status is high-dimensional and time-series, {directly using traditional data processing methods hardly abstracts accurate swipe probability distribution.} Therefore, we select \ac{gai} to process DT status data, which can effectively differentiate users' swipe behaviors for efficient resource management.

For performance comparison, we select two different schemes to substitute the proposed \ac{gdt} network decision-making module. Specifically, the heuristic scheme consists of the same GDT-assisted segment buffering strategy and different greedy resource scheduling, {while the \ac{drl} scheme includes the sequential segment buffering strategy and deep deterministic policy gradient-based resource scheduling.} Fig.~\ref{VM} shows the QoE trend with the increasing computing capacity. {It can be observed that the proposed data-model-driven method can always maintain the highest QoE, which reflects its ability to leverage additional computing resources to improve user experience. This is because the proposed data-model-driven method in the \ac{gdt} decision-making module can effectively abstract segment buffering information for tailored resource scheduling to improve \ac{qoe}. Furthermore, the heuristic scheme can always achieve better \ac{qoe} than \ac{drl} scheme, because \ac{gdt}-assisted segment buffering can efficiently abstract buffering information and the greedy resource scheduling can adapt well to different computing capacities.}


\section{Open Research Issues}

\subsection{Efficient GDT Module Collaboration}
Given the multi-modular composition of \ac{gdt} and the inherent interactivity among these modules, the efficient collaborative mechanism is critical for network performance enhancement. For instance, \ac{gai}-based feature abstraction module usually requires real-time and accurate network status from \ac{gai}-based status emulation module. {However, the data interaction frequency between the modules and data abstraction level in the GAI-based feature abstraction module can be adaptive and differentiated based on the network dynamics and service requirements.} {For instance, delay-tolerant tasks with low network dynamics usually require less data interaction between the modules and shallower feature extraction. Therefore, it is critical to develop an efficient \ac{gdt} module collaboration mechanism to reduce network cost.}

\subsection{Specialized Generative Model}
{Due to the substantial parameterization inherent in \ac{gai} models, the associated caching and computing overhead can significantly exacerbate network load. While deploying the models solely on cloud infrastructures may enable efficient model training and inference, it hardly provides satisfactory service for the task requiring low latency and high reliability. Consequently, {for delay-sensitive and high-reliability tasks, it is necessary to deploy lightweight and specialized models on network edge nodes. The approach also requires the fine-tuning of \ac{gai} models in response to network dynamics and service diversity}.}
\subsection{Efficient Resource Management for GDT Operation}

Although \ac{gdt} can efficiently process networking data to improve network management performance, the maintenance of \ac{gdt} is equally important, {which consumes substantial communication, caching, and computing resource.} {This is attributed to voluminous networking and user data storage, as well as resource-intensive \ac{gai} models for network status emulation, feature abstraction, and network decision-making.} {Since the limited network resources need to cater to both users' service requests and \ac{gdt} operation, it is imperative to accurately quantify the impact of \ac{gdt} operation on \ac{qos} and \ac{qoe}. {Based on the quantified result, the network resource scheduling for both \ac{gdt} operation and users' service requests can be more tailored to further improve \ac{qos} and \ac{qoe}.}}


\section{Conclusion}
\label{sec:Conclusion}

{We have proposed a \ac{gdt} network architecture to achieve intelligent external and internal closed-loop network management. Specifically, advanced \ac{gai} models are employed to improve \ac{dt} status emulation, feature abstraction, and network decision-making modules. In the external closed loop, \ac{gai}-based status emulation interacts with \ac{dt}-based error discriminator to adaptively adjust data collection frequency. In the internal closed loop, {\ac{gai}-based network decision-making algorithm collaboratives with model-based one in \ac{dt} to realize adaptive network management.} To further optimize \ac{gdt} network architecture, we have proposed a model light-weighting method, an adaptive model selection mechanism, and a data-model-driven method, respectively. A case study has been presented, and some open research issues have been discussed for accelerating the pace of \ac{gdt} network development.}




%
\bibliographystyle{IEEEtran}
\bibliography{IEEEabrv,Ref}
\end{document}